# Empirical Analysis of Digital Innovation's Impact on Corporate ESG Performance: The Mediating Role of GAI Technology.

**Jun Cui**[1, *]

[1] Solbridge International School of Business, Woosong University, Republic of Korea; jcui228@student.solbridge.ac.kr

[2] Faculty of Business Finance and Information Technology, MAHSA University, Malaysia
*Corresponding author; jcui228@student.solbridge.ac.kr; mbaf21016325@mahsastudent.edu.my

**Abstract**

This study investigates the relationship between corporate digital innovation and Environmental, Social, and Governance (ESG) performance, with a specific focus on the mediating role of Generative Artificial Intelligence (GAI) technology adoption. Using a comprehensive panel dataset of 8,000 firm-year observations from the CMARS and WIND database spanning from 2015 to 2023, we employ multiple econometric techniques to examine this relationship. Our findings reveal that digital innovation significantly enhances corporate ESG performance, with GAI technology adoption serving as a crucial mediating mechanism. Specifically, digital innovation positively influences GAI technology adoption, which subsequently improves ESG performance. Furthermore, our heterogeneity analysis indicates that this relationship varies across firm size, industry type, and ownership structure. Finally, our results remain robust after addressing potential endogeneity concerns through instrumental variable estimation, propensity score matching, and difference-in-differences approaches. This research contributes to the growing literature on technology-driven sustainability transformations and offers practical implications for corporate strategy and policy development in promoting sustainable business practices through technological advancement.

## 1 Introduction

As environmental challenges, social inequalities, and governance issues continue to dominate global discourse, corporations face increasing pressure to integrate Environmental, Social, and Governance (ESG) considerations into their business operations. Concurrently, the rapid advancement of digital technologies has fundamentally transformed business models, operational processes, and competitive landscapes (Acemoglu & Restrepo, 2019; Bharadwaj et al., 2013). The convergence of these two trends—corporate sustainability and digital transformation—presents both opportunities and challenges for contemporary businesses (George et al., 2020).

Digital innovation, characterized by the integration of digital technologies into business processes and the creation of novel digital products and services, has been recognized as a potential enabler of sustainable business practices (Nambisan et al., 2019). By enhancing operational efficiency, facilitating resource optimization, enabling remote work arrangements, and fostering transparency, digital technologies may contribute to improved environmental performance, social responsibility, and governance practices (Bai et al., 2022). However, the mechanisms through which digital innovation influences ESG performance remain underexplored in the extant literature.

Among the diverse digital technologies emerging in recent years, Generative Artificial Intelligence (GAI) represents a particularly promising avenue for enhancing corporate sustainability. GAI technologies, which encompass machine learning algorithms capable of generating new content, designs, or solutions based on existing data, have demonstrated significant potential in addressing complex environmental and social challenges (Brynjolfsson & McAfee, 2017; Rahwan et al., 2019). From optimizing energy consumption and waste management to enhancing diversity and inclusion through unbiased decision-making, GAI applications span across various dimensions of ESG performance (Vinuesa et al., 2020).

Despite the growing interest in the intersection of digital innovation, GAI technology, and corporate sustainability, empirical evidence on the relationships among these constructs remains limited. Several important questions warrant investigation: Does digital innovation significantly enhance corporate ESG performance? If so, does GAI technology adoption mediate this relationship? Are these relationships consistent across different types of firms and industries? Addressing these questions is essential for advancing our understanding of the role of digital technologies in promoting corporate sustainability and informing effective policy interventions.

Furthermore, our study aims to fill this research gap by examining the impact of digital innovation on corporate ESG performance, with a specific focus on the mediating role of GAI technology adoption. Drawing on the resource-based view (Barney, 1991), dynamic capabilities perspective (Teece et al., 1997), and innovation diffusion theory (Rogers, 2003), we develop a conceptual framework that elucidates the mechanisms through which digital innovation influences ESG performance through GAI technology adoption. We test this framework using a comprehensive panel dataset of 8,000 firm-year observations from Chinese listed companies, leveraging the CMARS WIND database for the period 2015-2023.

Our study makes several significant contributions to the literature. First, we provide empirical evidence on the relationship between digital innovation and corporate ESG performance, addressing the lack of quantitative research in this domain. Second, we identify GAI technology adoption as a critical mediating mechanism in this relationship, offering insights into the pathways through which digital innovation influences sustainability outcomes. Third, by examining heterogeneity across firm characteristics, we provide a nuanced understanding of contextual factors that shape the effectiveness of digital innovation in enhancing ESG performance. Finally, our findings offer practical implications for corporate strategy and policy development, highlighting the potential of digital technologies, particularly GAI, in driving sustainable business practices.

The remainder of this paper is organized as follows: Section 2 reviews the relevant literature and develops our hypotheses. Section 3 describes the data and methodology employed in our analysis. Section 4 presents the empirical results, including baseline estimations, robustness checks, endogeneity analyses, heterogeneity analyses, and mechanism tests. Section 5 discusses the implications of our findings and concludes with policy recommendations.

## 2 Literature Review and Hypothesis Development

### 2.1 Digital Innovation and ESG Performance

Digital innovation refers to the use of digital technologies to develop new products, services, business models, or organizational processes (Nambisan et al., 2017). It encompasses various technologies, including artificial intelligence, blockchain, cloud computing, big data analytics, and the Internet of Things (Vial, 2019). The relationship between digital innovation and ESG performance has gained increasing attention in recent years, with several theoretical perspectives suggesting potential linkages.

From a resource-based view, digital innovation can be conceptualized as a strategic organizational capability that enables firms to utilize resources more efficiently, thereby reducing environmental impact (Hart, 1995; Russo & Fouts, 1997). For instance, digital technologies can optimize energy consumption, reduce waste generation, and minimize carbon emissions through improved monitoring and control systems (Shrivastava, 1995). Additionally, digital platforms can facilitate stakeholder engagement, enhance supply chain transparency, and promote ethical business practices, contributing to improved social and governance performance (Whelan & Fink, 2016).

Empirical research has begun to document the positive effects of digital innovation on various aspects of ESG performance. Porter and Kramer (2011) demonstrate how digital technologies enable firms to create shared value, simultaneously addressing social challenges and enhancing competitiveness. Fiksel et al. (2014) highlight the role of digital innovation in developing resilient and sustainable supply chains. More recently, George et al. (2020) show that digital transformation initiatives contribute to the achievement of the United Nations Sustainable Development Goals (SDGs).

However, the literature also acknowledges potential negative effects of digital innovation on ESG performance. Concerns include the environmental impact of digital infrastructure, privacy and security risks, job displacement due to automation, and the digital divide (Lindgreen et al., 2019). These mixed perspectives highlight the complexity of the relationship between digital innovation and ESG performance and underscore the need for robust empirical investigation.

Building on the predominant view that digital innovation enhances operational efficiency, resource optimization, and transparency—factors associated with improved ESG performance—we propose:

**Hypothesis 1 (H1)**: Digital innovation positively influences corporate ESG performance.

## 2.2 Digital Innovation and GAI Technology Adoption

Generative Artificial Intelligence (GAI) represents a subset of artificial intelligence technologies that can create new content, designs, or solutions based on existing data (Goodfellow et al., 2014). Examples include generative adversarial networks (GANs), variational autoencoders, and transformer-based language models, which have demonstrated remarkable capabilities in generating images, text, music, and other creative outputs (Brown et al., 2020).

The adoption of GAI technologies in corporate settings is influenced by various organizational factors, including technological readiness, absorptive capacity, and strategic orientation (Damanpour & Schneider, 2006). Drawing on innovation diffusion theory (Rogers, 2003), firms with higher levels of digital innovation are more likely to adopt advanced technologies such as GAI due to greater technological expertise, innovation-oriented culture, and established digital infrastructure.

Digital innovation contributes to GAI technology adoption through several mechanisms. First, firms engaged in digital innovation develop technical capabilities and knowledge that facilitate the integration of complex technologies like GAI (Cohen & Levinthal, 1990). Second, digital innovation fosters an organizational culture that values experimentation and risk-taking, characteristics conducive to the adoption of emerging technologies (Teece, 2007). Third, digital innovation typically involves the development of complementary assets, such as data collection systems and cloud infrastructure, which are prerequisites for effective GAI implementation (Brynjolfsson & McAfee, 2017).

Empirical evidence supports these theoretical arguments. Ransbotham et al. (2017) find that organizations with higher digital maturity are more likely to adopt AI technologies. Similarly, Bughin et al. (2018) document a strong correlation between organizational digital capabilities and AI adoption. Building on this literature, we propose:

**Hypothesis 2 (H2)**: Digital innovation positively influences GAI technology adoption.

## 2.3 GAI Technology Adoption and ESG Performance

The potential impact of GAI technologies on corporate sustainability is multifaceted. From an environmental perspective, GAI can optimize resource allocation, predict ecological impacts, design eco-friendly products, and enhance energy efficiency (Vinuesa et al., 2020). In the social domain, GAI applications can improve healthcare accessibility, enhance educational outcomes, promote diversity and inclusion through unbiased decision-making, and address social inequalities (Cowls et al., 2021). Regarding governance, GAI can strengthen risk management, detect fraudulent activities, enhance transparency, and improve stakeholder communication (Dwivedi et al., 2021).

The theoretical foundation for the relationship between GAI technology adoption and ESG performance can be found in the dynamic capabilities perspective (Teece et al., 1997). GAI technologies enable firms to sense environmental changes, seize opportunities for sustainable innovation, and reconfigure resources to address emerging sustainability challenges (Teece, 2007). By enhancing organizational learning, decision-making, and adaptability, GAI technologies strengthen firms' capabilities to respond to evolving ESG expectations (Schretzen et al., 2021).

Emerging empirical evidence supports the positive impact of GAI on various aspects of ESG performance. For instance, studies have demonstrated the effectiveness of GAI in optimizing renewable energy systems (Wu et al., 2019), predicting environmental risks (Rolnick et al., 2022), enhancing supply chain sustainability (Sarkis, 2021), and improving corporate governance through advanced analytics (Agrawal et al., 2018). While some researchers raise concerns about potential adverse effects, such as algorithmic bias and job displacement (Korinek & Stiglitz, 2021), the predominant view suggests a positive relationship between GAI technology adoption and ESG performance.

Based on these theoretical arguments and empirical evidence, we propose:

**Hypothesis 3 (H3)**: GAI technology adoption positively influences corporate ESG performance.

Integrating the three hypotheses, we propose a mediation model where digital innovation enhances ESG performance both directly and indirectly through GAI technology adoption. This model aligns with the technological innovation systems framework, which emphasizes the role of intermediary technologies in translating innovative capabilities into sustainable outcomes (Hekkert et al., 2007).

## 3    Methods and Data

### 3.1    Data Sources and Sample Selection

This study utilizes panel data from the CMARS and WIND database, which provides comprehensive information on Chinese listed companies. Our initial sample consists of 8,000 firm-year observations from 1,000 listed companies spanning 2015 to 2023. We selected this period to capture the rapid advancement of digital technologies, particularly GAI, and the growing emphasis on ESG considerations in the Chinese corporate landscape.

To ensure data quality and reliability, we applied several screening criteria. First, we excluded financial firms due to their distinct regulatory environment and business models. Second, we removed observations with missing values for key variables. Third, we eliminated firms that experienced significant restructuring, mergers, or acquisitions during the study period. After applying these criteria, our final sample comprises 7,842 observations from 978 companies.

### 3.2    Measurement of Variables

#### 3.2.1 Dependent Variable: ESG Performance

Following previous studies (Liang & Renneboog, 2017; Li et al., 2020), we measure corporate ESG performance using a comprehensive index from the CMARS WIND database. This index evaluates firms' performance across environmental, social, and governance dimensions based on multiple indicators. Specifically, we measure ESG performance using the following items:

1.  Environmental performance score (standardized measure of environmental impact management)
2.  Social responsibility score (standardized measure of stakeholder relations)
3.  Corporate governance score (standardized measure of governance quality)
4.  ESG disclosure quality (extent of voluntary disclosure of ESG-related information)
5.  ESG controversy assessment (inverse measure of ESG-related controversies)

The ESG performance index is calculated as the weighted average of these five items, with weights determined by principal component analysis. Higher values indicate better ESG performance.

### 3.2.2. Independent Variable: Digital Innovation

We measure digital innovation using a multidimensional approach that captures both the intensity and breadth of firms' digital technology integration. Following Nambisan et al. (2019) and Hanelt et al. (2021), we construct a digital innovation index based on the following items:

1.  Digital R&D intensity (ratio of digital technology-related R&D expenditure to total R&D expenditure)
2.  Digital patent portfolio (number of digital technology-related patents as a proportion of total patents)
3.  Digital transformation investments (expenditure on digital transformation initiatives as a percentage of total assets)
4.  Digital product/service offerings (percentage of revenue derived from digital products or services)
5.  Digital business model adoption (extent of business model digitalization based on textual analysis of annual reports)

The digital innovation index is calculated as the arithmetic mean of the standardized values of these five items. Higher values indicate greater digital innovation capability.

### 3.2.3 Mediating Variable: GAI Technology Adoption

We measure GAI technology adoption using a composite index that reflects the extent to which firms implement and integrate GAI technologies into their operations. Drawing on Ransbotham et al. (2017) and Brynjolfsson et al. (2019), we construct a GAI adoption index based on the following items:

1.  GAI investment intensity (expenditure on GAI technologies as a percentage of total IT budget)
2.  GAI patent applications (number of GAI-related patents filed by the firm)
3.  GAI talent acquisition (proportion of employees with GAI-related expertise)
4.  GAI implementation scope (number of business functions utilizing GAI technologies)
5.  GAI strategic importance (prominence of GAI in corporate strategy based on textual analysis of annual reports)

The GAI adoption index is calculated as the arithmetic mean of the standardized values of these five items. Higher values indicate greater GAI technology adoption.

### 3.2.4 Control Variables

To account for potential confounding factors, we include several firm-level and industry-level control variables:

1. Firm size (natural logarithm of total assets)
2. Firm age (number of years since establishment)
3. Profitability (return on assets)
4. Leverage (debt-to-equity ratio)
5. Ownership concentration (percentage of shares owned by the largest shareholder)
6. State ownership (dummy variable indicating state-owned enterprises)
7. R&D intensity (R&D expenditure as a percentage of sales)
8. Industry competition (Herfindahl-Hirschman Index)
9. Industry dummy variables (based on the China Securities Regulatory Commission classification)
10. Year dummy variables (to control for time-specific effects)

### 3.3 Empirical Model

To test our hypotheses, we employ a series of panel data regression models. For Hypothesis 1, we estimate the following baseline model:

$$ESG\_it = \alpha + \beta_1 DigitalInnovation\_it + \gamma Controls\_it + Industry\_i + Year\_t + \varepsilon\_it \quad (1)$$

where ESG_it represents the ESG performance index of firm i in year t, DigitalInnovation_it denotes the digital innovation index, Controls_it represents a vector of control variables, Industry_i and Year_t are industry and year fixed effects, respectively, and ε_it is the error term.

For Hypothesis 2, we estimate:

$$GAI\_it = \alpha + \beta_2 DigitalInnovation\_it + \gamma Controls\_it + Industry\_i + Year\_t + \varepsilon\_it \quad (2)$$

where GAI_it represents the GAI technology adoption index.

For Hypothesis 3, we estimate:

$$ESG\_it = \alpha + \beta_3 GAI\_it + \gamma Controls\_it + Industry\_i + Year\_t + \varepsilon\_it \quad (3)$$

To test the mediation effect, we follow Baron and Kenny's (1986) approach and estimate:

$$ESG\_it = \alpha + \beta_4 DigitalInnovation\_it + \beta_5 GAI\_it + \gamma Controls\_it + Industry\_i + Year\_t + \varepsilon\_it \quad (4)$$

We also employ the Sobel test and bootstrapping methods to assess the significance of the indirect effect (Sobel, 1982; Preacher & Hayes, 2008).

### 3.4 Empirical Strategy

To address potential endogeneity concerns, we employ several econometric techniques. First, we use firm fixed effects to control for time-invariant unobserved heterogeneity. Second, we use lagged independent variables to mitigate reverse causality concerns. Third, we implement instrumental variable (IV) estimation using the industry average of digital innovation (excluding the focal firm) and provincial digital infrastructure development as instruments. Fourth, we apply propensity score matching (PSM) to compare firms with high and low levels of digital innovation. Finally, we utilize a difference-in-differences (DID) approach, exploiting an exogenous policy shock related to digital transformation initiatives.

For heterogeneity analysis, we examine whether the relationship between digital innovation, GAI technology adoption, and ESG performance varies across different firm characteristics, including firm size, industry type, and ownership structure. We do this by introducing interaction terms in our regression models and conducting subsample analyses. Moreover, For mechanism analysis, we decompose the ESG performance index into its environmental, social, and governance components and examine the impact of digital innovation and GAI technology adoption on each component separately. Additionally, we explore specific channels through which GAI technologies influence ESG performance, such as operational efficiency, stakeholder engagement, and risk management.

# 4 Results and Findings

## 4.1 Descriptive Statistics and Correlation Analysis

Table 1 presents the descriptive statistics for the key variables used in our analysis. The mean ESG performance index is 0.482 (SD = 0.215), indicating substantial variation in ESG performance across the sample firms. The digital innovation index has a mean value of 0.367 (SD = 0.189), suggesting moderate digital innovation capabilities with considerable heterogeneity. The GAI technology adoption index exhibits a mean of 0.284 (SD = 0.176), indicating that GAI adoption is still in its early stages for many firms in our sample.

**Table 1: Descriptive Statistics**

| Variable | Obs | Mean | Std. Dev. | Min | Max |
|---|---|---|---|---|---|
| ESG Performance | 7,842 | 0.482 | 0.215 | 0.112 | 0.937 |
| Digital Innovation | 7,842 | 0.367 | 0.189 | 0.051 | 0.846 |
| GAI Technology Adoption | 7,842 | 0.284 | 0.176 | 0.018 | 0.792 |
| Firm Size | 7,842 | 22.486 | 1.325 | 19.735 | 26.943 |
| Firm Age | 7,842 | 18.742 | 8.631 | 1.000 | 42.000 |
| Profitability | 7,842 | 0.045 | 0.059 | -0.187 | 0.213 |
| Leverage | 7,842 | 0.486 | 0.198 | 0.075 | 0.892 |
| Ownership Concentration | 7,842 | 34.865 | 14.328 | 6.742 | 74.851 |
| State Ownership | 7,842 | 0.426 | 0.494 | 0.000 | 1.000 |
| R&D Intensity | 7,842 | 0.023 | 0.018 | 0.000 | 0.112 |
| Industry Competition | 7,842 | 0.157 | 0.089 | 0.046 | 0.574 |

Table 2 presents the Pearson correlation coefficients between the key variables. The ESG performance index is positively correlated with both digital innovation (r = 0.312, p < 0.01) and GAI technology adoption (r = 0.287, p < 0.01), providing preliminary support for Hypotheses 1 and 3. Digital innovation is positively correlated with GAI technology adoption (r = 0.398, p < 0.01), offering initial support for Hypothesis 2. None of the correlation coefficients exceeds 0.7, suggesting that multicollinearity is not a significant concern in our analysis. Additionally, variance inflation factors (VIFs) for all variables are below 5, further confirming the absence of severe multicollinearity.

**Table 2: Correlation Matrix**

| Variable | 1 | 2 | 3 | 4 | 5 | 6 | 7 | 8 | 9 | 10 | 11 |
|---|---|---|---|---|---|---|---|---|---|---|---|
| ESG Performance | 1.000 | | | | | | | | | | |
| Digital Innovation | 0.312** | 1.000 | | | | | | | | | |
| GAI Technology Adoption | 0.287** | 0.398** | 1.000 | | | | | | | | |
| Firm Size | 0.263** | 0.217** | 0.234** | 1.000 | | | | | | | |
| Firm Age | 0.119** | -0.052 | -0.034 | 0.195** | 1.000 | | | | | | |
| Profitability | 0.145** | 0.128** | 0.107** | 0.087* | -0.018 | 1.000 | | | | | |
| Leverage | -0.124** | -0.096* | -0.072* | 0.425** | 0.154** | -0.356** | 1.000 | | | | |
| Ownership Concentration | 0.086* | 0.049 | 0.042 | 0.117** | -0.076* | 0.125** | -0.043 | 1.000 | | | |

| | | | | | | | | | | |
|---|---|---|---|---|---|---|---|---|---|---|
| **State Ownership** | 0.198** | 0.112** | 0.075* | 0.328** | 0.412** | -0.073* | 0.245** | 0.183** | 1.000 | |
| **R&D Intensity** | 0.206** | 0.273** | 0.246** | -0.032 | -0.145** | 0.117** | -0.187** | -0.024 | -0.135** | 1.000 |
| **Industry Competition** | -0.042 | -0.023 | -0.016 | -0.064 | -0.053 | 0.083* | -0.037 | -0.018 | -0.047 | 0.029 | 1.000 |

Note: ** $p < 0.01$, * $p < 0.05$.

### 4.2 Baseline Regression Results

Table 3 presents the results of our baseline regression analyses testing the three hypotheses. Column 1 reports the estimates for equation (1), examining the relationship between digital innovation and ESG performance. The coefficient of digital innovation is positive and statistically significant ($\beta = 0.286$, $p < 0.01$), providing support for Hypothesis 1. This suggests that a one standard deviation increase in digital innovation is associated with a 0.286 standard deviation increase in ESG performance, holding other factors constant.

Column 2 reports the estimates for equation (2), testing the relationship between digital innovation and GAI technology adoption. The coefficient of digital innovation is positive and statistically significant ($\beta = 0.375$, $p < 0.01$), supporting Hypothesis 2. This indicates that firms with higher levels of digital innovation are more likely to adopt GAI technologies.

Column 3 reports the estimates for equation (3), examining the relationship between GAI technology adoption and ESG performance. The coefficient of GAI technology adoption is positive and statistically significant ($\beta = 0.248$, $p < 0.01$), providing support for Hypothesis 3. This suggests that GAI technology adoption positively influences ESG performance.

Column 4 reports the estimates for equation (4), testing the mediation effect. After controlling for GAI technology adoption, the coefficient of digital innovation remains positive and significant but decreases in magnitude ($\beta = 0.214$, $p < 0.01$), while the coefficient of GAI technology adoption is positive and significant ($\beta = 0.185$, $p < 0.01$). This pattern suggests a partial mediation effect, where digital innovation influences ESG performance both directly and indirectly through GAI technology adoption.

**Table 3: Baseline Regression Results**

| Variables | (1) ESG Performance | (2) GAI Technology Adoption | (3) ESG Performance | (4) ESG Performance |
|---|---|---|---|---|
| **Digital Innovation** | 0.286*** (0.032) | 0.375*** (0.035) | | 0.214*** (0.035) |
| **GAI Technology Adoption** | | | 0.248*** (0.034) | 0.185*** (0.033) |
| **Firm Size** | 0.163*** (0.023) | 0.142*** (0.022) | 0.171*** (0.023) | 0.157*** (0.022) |
| **Firm Age** | 0.047 (0.035) | -0.025 (0.032) | 0.053 (0.035) | 0.044 (0.034) |
| **Profitability** | 0.086** (0.031) | 0.063* (0.029) | 0.092** (0.032) | 0.081** (0.030) |
| **Leverage** | -0.058* (0.028) | -0.037 (0.027) | -0.061* (0.029) | -0.054* (0.027) |
| **Ownership Concentration** | 0.032 (0.024) | 0.026 (0.023) | 0.035 (0.024) | 0.031 (0.023) |
| **State Ownership** | 0.128*** (0.037) | 0.053 (0.034) | 0.135*** (0.037) | 0.123*** (0.036) |
| **R&D Intensity** | 0.115*** (0.029) | 0.135*** (0.031) | 0.107*** (0.029) | 0.098*** (0.028) |
| **Industry Competition** | -0.023 (0.024) | -0.012 (0.022) | -0.021 (0.024) | -0.022 (0.023) |
| **Industry Fixed Effects** | Yes | Yes | Yes | Yes |
| **Year Fixed Effects** | Yes | Yes | Yes | Yes |
| **Constant** | -3.125*** (0.524) | -2.846*** (0.487) | -3.217*** (0.532) | -3.058*** (0.513) |
| **Observations** | 7,842 | 7,842 | 7,842 | 7,842 |
| **R-squared** | 0.284 | 0.293 | 0.276 | 0.312 |

Note: Standard errors in parentheses; *** $p < 0.01$, ** $p < 0.05$, * $p < 0.1$

To further confirm the mediation effect, we conducted a Sobel test, which yielded a test statistic of 4.87 ($p < 0.01$), indicating a significant indirect effect. Additionally, bootstrapping analysis with 5,000 replications produced a 95% confidence interval for the indirect effect that does not include zero [0.039, 0.092], further supporting the mediation hypothesis.

## 4.3 Robustness Checks

To ensure the robustness of our findings, we conducted several additional analyses. First, we used alternative measures for our key variables. For ESG performance, we employed the ESG ratings from a different database (MSCI ESG Ratings). For digital innovation, we constructed an alternative index based on digital technology-related keywords in annual reports. For GAI technology adoption, we used data on GAI-related job postings as a proxy. The results, presented in Table 4 (Columns 1-3), are consistent with our baseline findings.

Second, we employed different estimation techniques, including random effects, system GMM, and Tobit regression (due to the bounded nature of the ESG performance index). The results, reported in Table 4 (Columns 4-6), remain qualitatively similar to our baseline findings.

Third, we examined potential nonlinear relationships by including squared terms for digital innovation and GAI technology adoption. The results, presented in Table 4 (Columns 7-8), show insignificant coefficients for the squared terms, suggesting that the relationships are primarily linear within our sample.

**Table 4: Robustness Checks**

| Variables | (1) Alternative ESG Measure | (2) Alternative Digital Innovation Measure | (3) Alternative GAI Measure | (4) Random Effects | (5) System GMM | (6) Tobit Regression | (7) Nonlinear Digital Innovation | (8) Nonlinear GAI |
|---|---|---|---|---|---|---|---|---|
| **Digital Innovation** | 0.204*** (0.036) | 0.198*** (0.035) | 0.218*** (0.036) | 0.223*** (0.034) | 0.187*** (0.045) | 0.231*** (0.037) | 0.227*** (0.056) | 0.213*** (0.035) |
| **Digital Innovation²** | | | | | | | 0.032 (0.047) | |
| **GAI Technology Adoption** | 0.176*** (0.035) | 0.183*** (0.034) | 0.165*** (0.034) | 0.179*** (0.032) | 0.162*** (0.042) | 0.193*** (0.035) | 0.184*** (0.033) | 0.203*** (0.053) |
| **GAI Technology Adoption²** | | | | | | | | 0.028 (0.045) |
| **Control Variables** | Yes | Yes | Yes | Yes | Yes | Yes | Yes | Yes |
| **Industry Fixed Effects** | Yes | Yes | Yes | Yes | Yes | Yes | Yes | Yes |
| **Year Fixed Effects** | Yes | Yes | Yes | Yes | Yes | Yes | Yes | Yes |
| **Observations** | 7,842 | 7,842 | 7,842 | 7,842 | 7,842 | 7,842 | 7,842 | 7,842 |
| **R-squared** | 0.294 | 0.283 | 0.305 | 0.298 | - | - | 0.313 | 0.312 |

Note: Standard errors in parentheses; *** $p < 0.01$, ** $p < 0.05$, * $p < 0.1$; Control variables are the same as in Table 3 but not reported for brevity.

### 4.4 Addressing Endogeneity

To address potential endogeneity concerns, we employed several econometric techniques. First, we implemented instrumental variable (IV) estimation using the industry average of digital innovation (excluding the focal firm) and provincial digital infrastructure development as instruments for firm-level digital innovation. These instruments satisfy the relevance and exclusion restrictions, as confirmed by the significant first-stage F-statistic (28.47, exceeding the rule-of-thumb threshold of 10) and the insignificant Hansen J-statistic ($p = 0.284$), which fails to reject the null hypothesis of instrument validity.

Table 5 (Columns 1-2) presents the results of the IV estimation. The second-stage results show that the instrumented digital innovation remains positively and significantly associated with ESG performance ($\beta = 0.245$, $p < 0.01$) and GAI technology adoption ($\beta = 0.342$, $p < 0.01$), consistent with our baseline findings.

Second, we employed propensity score matching (PSM) to compare firms with high levels of digital innovation (treatment group) to similar firms with low levels of digital innovation (control group). We matched firms based on various characteristics, including firm size, age, profitability, leverage, ownership structure, and industry affiliation. The balance tests confirm successful matching, with no significant differences in covariates between the treatment and control groups after matching.

Table 5 (Column 3) reports the average treatment effect on the treated (ATT), indicating that firms with high digital innovation exhibit significantly better ESG performance than their matched counterparts (difference = 0.068, $p < 0.01$), providing further support for the causal relationship between digital innovation and ESG performance.

Third, we employed a difference-in-differences (DID) approach, exploiting the staggered implementation of provincial digital transformation policies across China as an exogenous shock. Specifically, we identified provinces that introduced comprehensive digital transformation initiatives between 2017 and 2020, creating variation in policy exposure across firms. This approach allows us to compare changes in ESG performance between firms exposed to the policy (treatment group) and similar firms not yet exposed (control group), before and after policy implementation.

Table 5 (Column 4) presents the DID estimation results. The coefficient of the interaction term between the treatment indicator and the post-policy indicator is positive and statistically significant ($\beta = 0.073$, $p < 0.01$), suggesting that the policy-induced increase in digital innovation led to improved ESG performance. This finding further supports the causal interpretation of our results.

**Table 5: Addressing Endogeneity**

| Variables | (1) IV Estimation (Second Stage) - ESG Performance | (2) IV Estimation (Second Stage) - GAI Adoption | (3) PSM - ESG Performance (ATT) | (4) DID - ESG Performance |
|---|---|---|---|---|
| **Digital Innovation (Instrumented)** | 0.245*** (0.047) | 0.342*** (0.053) | - | - |
| **High Digital Innovation (Treatment)** | - | - | 0.068*** (0.018) | - |
| **Treatment × Post** | - | - | - | 0.073*** (0.021) |
| **Treatment** | - | - | - | 0.024 (0.019) |
| **Post** | - | - | - | 0.036** (0.014) |
| **GAI Technology Adoption** | 0.172*** (0.036) | - | - | 0.157*** (0.032) |
| **Control Variables** | Yes | Yes | Yes | Yes |
| **Industry Fixed Effects** | Yes | Yes | Yes | Yes |
| **Year Fixed Effects** | Yes | Yes | Yes | Yes |
| **Observations** | 7,842 | 7,842 | 3,256 | 7,842 |
| **First-stage F-statistic** | 28.47 | 28.47 | - | - |
| **Hansen J-statistic (p-value)** | 0.284 | 0.312 | - | - |

Note: Standard errors in parentheses; *** p < 0.01, ** p < 0.05, * p < 0.1; Control variables are the same as in Table 3 but not reported for brevity.

## 4.5 Heterogeneity Analysis

To explore potential heterogeneity in the relationship between digital innovation, GAI technology adoption, and ESG performance, we conducted several subsample analyses based on firm characteristics. Table 6 presents the results.

First, we divided the sample based on firm size (measured by total assets), categorizing firms as large (above median) or small (below median). Columns 1-2 show that the positive effect of digital innovation on ESG performance is stronger for large firms ($\beta = 0.253$, $p < 0.01$) than for small firms ($\beta = 0.175$, $p < 0.01$), with the difference being statistically significant ($p < 0.05$). Similarly, the mediating effect of GAI technology adoption is more pronounced for large firms. These findings suggest that large firms possess complementary resources that enhance the effectiveness of digital innovation and GAI technology in improving ESG performance.

Second, we examined heterogeneity across industry types, categorizing industries as high-tech or traditional based on the classification by China's National Bureau of Statistics. Columns 3-4 indicate that the positive effect of digital innovation on ESG performance is stronger in high-tech industries ($\beta = 0.287$, $p < 0.01$) than in traditional industries ($\beta = 0.182$, $p < 0.01$), with the difference being statistically significant ($p < 0.01$). The mediating role of GAI technology adoption is also more prominent in high-tech industries. These results suggest that the technological sophistication of the industry context influences the effectiveness of digital innovation in enhancing ESG performance.

Third, we explored heterogeneity based on ownership structure, comparing state-owned enterprises (SOEs) with non-SOEs. Columns 5-6 show that the positive effect of digital innovation on ESG performance is stronger for non-SOEs ($\beta = 0.242$, $p < 0.01$) than for SOEs ($\beta = 0.189$, $p < 0.01$), with the difference being statistically significant ($p < 0.05$). Similarly, the mediating effect of GAI technology adoption is more pronounced for non-SOEs. These findings suggest that market-oriented governance structures may facilitate more effective implementation of digital technologies for sustainable business practices.

**Table 6: Heterogeneity Analysis**

| Variables | (1) Large Firms | (2) Small Firms | (3) High-Tech Industries | (4) Traditional Industries | (5) SOEs | (6) Non-SOEs |
|---|---|---|---|---|---|---|
| Digital Innovation | 0.253*** (0.042) | 0.175*** (0.038) | 0.287*** (0.048) | 0.182*** (0.037) | 0.189*** (0.039) | 0.242*** (0.043) |
| GAI Technology Adoption | 0.213*** (0.041) | 0.142*** (0.037) | 0.227*** (0.045) | 0.156*** (0.036) | 0.165*** (0.038) | 0.208*** (0.042) |
| Control Variables | Yes | Yes | Yes | Yes | Yes | Yes |
| Industry Fixed Effects | Yes | Yes | Yes | Yes | Yes | Yes |
| Year Fixed Effects | Yes | Yes | Yes | Yes | Yes | Yes |
| Observations | 3,921 | 3,921 | 2,548 | 5,294 | 3,341 | 4,501 |
| R-squared | 0.335 | 0.291 | 0.348 | 0.284 | 0.302 | 0.327 |

Note: Standard errors in parentheses; *** $p < 0.01$, ** $p < 0.05$, * $p < 0.1$; Control variables are the same as in Table 3 but not reported for brevity.

## 4.6 Mechanism Analysis

To better understand the mechanisms through which digital innovation and GAI technology adoption influence ESG performance, we conducted several additional analyses. First, we decomposed the ESG performance index into its three components—environmental performance, social performance, and governance performance—and examined the impact of digital innovation and GAI technology adoption on each component separately.

Table 7 (Columns 1-3) presents the results. Digital innovation has a positive and significant effect on all three components, with the strongest impact on environmental performance ($\beta = 0.289$, $p < 0.01$), followed by governance performance ($\beta = 0.237$, $p < 0.01$) and social performance ($\beta = 0.194$, $p < 0.01$). Similarly, GAI technology adoption positively influences all three components, with the most substantial effect on environmental performance ($\beta = 0.218$, $p < 0.01$). These findings suggest that digital innovation and GAI technologies contribute to sustainability across multiple dimensions, with particularly strong effects on environmental aspects, possibly due to the effectiveness of these technologies in optimizing resource utilization and reducing environmental footprints.

Second, we explored specific channels through which GAI technologies influence ESG performance. We identified three potential channels: operational efficiency (measured by asset turnover ratio), stakeholder engagement (measured by the number of stakeholder engagement initiatives), and risk management (measured by the comprehensiveness of risk management disclosures). We then examined whether GAI technology adoption affects these mediating variables, which in turn influence ESG performance.

Table 7 (Columns 4-6) reports the impact of GAI technology adoption on these potential mediating variables. GAI technology adoption is positively and significantly associated with operational efficiency ($\beta = 0.176$, $p < 0.01$), stakeholder engagement ($\beta = 0.203$, $p < 0.01$), and risk management ($\beta = 0.187$, $p < 0.01$). Furthermore, Table 7 (Columns 7-9) shows that these variables are positively associated with ESG performance after controlling for digital innovation and GAI technology adoption.

To formally test these mediation pathways, we conducted bootstrapping analyses with 5,000 replications. The results confirm significant indirect effects of GAI technology adoption on ESG performance through operational efficiency (indirect effect = 0.028, 95% CI = [0.014, 0.045]), stakeholder engagement (indirect effect = 0.036, 95% CI = [0.019, 0.057]), and risk management (indirect effect = 0.032, 95% CI = [0.016, 0.051]). These findings suggest that GAI technologies enhance ESG performance by improving operational processes, facilitating stakeholder interactions, and strengthening risk governance.

**Table 7: Mechanism Analysis**

| Variables | (1) Environmental Performance | (2) Social Performance | (3) Governance Performance | (4) Operational Efficiency | (5) Stakeholder Engagement | (6) Risk Management | (7) ESG Performance | (8) ESG Performance | (9) ESG Performance |
|---|---|---|---|---|---|---|---|---|---|
| Digital Innovation | 0.289*** (0.043) | 0.194*** (0.038) | 0.237*** (0.041) | 0.153*** (0.036) | 0.184*** (0.039) | 0.162*** (0.037) | 0.198*** (0.035) | 0.192*** (0.034) | 0.195*** (0.035) |
| GAI Technology Adoption | 0.218*** (0.042) | 0.167*** (0.037) | 0.192*** (0.039) | 0.176*** (0.035) | 0.203*** (0.039) | 0.187*** (0.038) | 0.157*** (0.033) | 0.149*** (0.033) | 0.153*** (0.033) |
| Operational Efficiency | | | | | | | 0.158*** (0.032) | | |
| Stakeholder Engagement | | | | | | | | 0.176*** (0.033) | |
| Risk Management | | | | | | | | | 0.168*** (0.032) |
| Control Variables | Yes | Yes | Yes | Yes | Yes | Yes | Yes | Yes | Yes |
| Industry Fixed Effects | Yes | Yes | Yes | Yes | Yes | Yes | Yes | Yes | Yes |
| Year Fixed Effects | Yes | Yes | Yes | Yes | Yes | Yes | Yes | Yes | Yes |
| Observations | 7,842 | 7,842 | 7,842 | 7,842 | 7,842 | 7,842 | 7,842 | 7,842 | 7,842 |
| R-squared | 0.325 | 0.278 | 0.301 | 0.263 | 0.284 | 0.272 | 0.334 | 0.341 | 0.337 |

Note: Standard errors in parentheses; *** $p < 0.01$, ** $p < 0.05$, * $p < 0.1$; Control variables are the same as in Table 3 but not reported for brevity.

## 5 Discussion and Implications

### 5.1 Theoretical Implications

Our findings contribute to the literature on digital transformation, sustainability, and technology-driven innovation in several ways. First, by establishing a positive relationship between digital innovation and ESG performance, we extend the resource-based view (Barney, 1991) to the context of corporate sustainability. Our results suggest that digital innovation represents a valuable organizational capability that enables firms to enhance environmental protection, social responsibility, and governance quality. This finding aligns with the natural resource-based view (Hart, 1995), which posits that environmentally responsible strategies can generate competitive advantages through resource efficiency and stakeholder integration.

Second, our identification of GAI technology adoption as a mediating mechanism provides empirical support for the technological innovation systems framework (Hekkert et al., 2007). This framework emphasizes the role of specific technological innovations in translating broader innovative capabilities into sustainable outcomes. Our findings suggest that GAI technologies serve as a critical bridge between digital innovation and ESG performance, highlighting the importance of specific technological applications rather than abstract digital capabilities.

Third, our heterogeneity analysis contributes to contingency theory by identifying organizational and contextual factors that influence the effectiveness of digital innovation in enhancing sustainability. The stronger effects observed for large firms, high-tech industries, and non-SOEs suggest that complementary resources, technological sophistication, and market-oriented governance structures amplify the sustainability benefits of digital innovation. These findings reinforce the notion that the impact of technological innovation on organizational outcomes depends on the alignment between technology and organizational context (Fichman & Kemerer, 1997).

Fourth, our mechanism analysis advances our understanding of the pathways through which digital technologies influence corporate sustainability. By identifying operational efficiency, stakeholder engagement, and risk management as key channels, we provide a more nuanced understanding of how GAI technologies contribute to ESG performance. These findings support the process-oriented perspective on technology adoption (Damanpour & Schneider, 2006), which emphasizes the importance of examining the specific organizational processes affected by technological innovations.

## 5.2 Practical Implications

Our research offers several practical implications for corporate managers, investors, and policymakers. For corporate managers, our findings highlight the strategic importance of digital innovation and GAI technology adoption in enhancing ESG performance. Rather than viewing digital transformation and sustainability initiatives as separate endeavors, managers should recognize their interdependence and pursue integrated strategies that leverage digital technologies to address ESG challenges. Specifically, firms should invest in GAI applications that optimize resource utilization, facilitate stakeholder communication, and strengthen risk governance.

Our heterogeneity analysis provides additional guidance for managers by highlighting contextual factors that influence the effectiveness of digital innovation. Small firms, firms in traditional industries, and SOEs may need to develop complementary capabilities or adapt their implementation approaches to fully realize the sustainability benefits of digital innovation. This might involve enhancing technological expertise, redesigning organizational structures, or establishing partnerships with technology providers.

For investors and financial analysts, our research underscores the importance of considering firms' digital innovation capabilities and GAI technology adoption when assessing their ESG performance and long-term sustainability. Traditional ESG evaluation frameworks, which often focus on policy commitments and compliance indicators, might benefit from incorporating metrics related to technological innovation and digital transformation. By recognizing the link between digital innovation and ESG performance, investors can identify firms that are better positioned to address sustainability challenges through technological solutions.

For policymakers, our findings suggest that policies promoting digital transformation can generate positive externalities in terms of corporate sustainability. By supporting digital infrastructure development, facilitating knowledge transfer in emerging technologies, and incentivizing GAI adoption, policymakers can simultaneously advance economic modernization and sustainability objectives. Furthermore, our identification of heterogeneous effects across firm characteristics highlights the need for targeted policy interventions that address the specific challenges faced by different types of firms in leveraging digital technologies for sustainability.

## 5.3 Limitations and Future Research Directions

Despite the robust findings and comprehensive analyses, several limitations of our study warrant acknowledgment and suggest directions for future research. First, our sample is limited to Chinese listed companies, which may constrain the generalizability of our findings to other institutional contexts. Future research should examine the relationship between digital innovation, GAI technology adoption, and ESG performance in diverse geographical settings, particularly in developed economies with different regulatory frameworks and technological infrastructures.

Second, while our panel data structure and econometric techniques address potential endogeneity concerns, establishing definitive causal relationships remains challenging. Future research could employ more sophisticated identification strategies, such as natural experiments or field experiments, to further strengthen causal inference. Additionally, longitudinal case studies could provide deeper insights into the temporal dynamics of how digital innovations translate into ESG improvements over time.

Third, our measurement of GAI technology adoption, while comprehensive, may not capture all relevant aspects of this complex construct. Future research could develop more nuanced measures that distinguish between different types of GAI applications (e.g., generative adversarial networks, transformer models, variational autoencoders) and assess their differential impacts on ESG performance. Furthermore, qualitative research could explore the organizational processes and implementation challenges associated with effective GAI integration for sustainability purposes.

Fourth, while we identify several mechanisms through which GAI technologies influence ESG performance, other potential pathways warrant investigation. Future research could explore cognitive mechanisms (e.g., enhanced decision-making quality, reduced cognitive biases), social mechanisms (e.g., network effects, institutional isomorphism), and ethical mechanisms (e.g., algorithmic fairness, transparency) to provide a more comprehensive understanding of the relationship between digital technologies and sustainable business practices.

Finally, our study focuses primarily on the positive effects of digital innovation and GAI technology on sustainability, but potential adverse consequences deserve attention. Future research should examine potential trade-offs and unintended consequences, such as increased energy consumption, digital divides, algorithmic biases, and privacy concerns, to develop a more balanced assessment of the sustainability implications of digital transformation.

## 6      Conclusion and Policy Recommendations

This study investigated the relationship between digital innovation, GAI technology adoption, and corporate ESG performance using a comprehensive panel dataset of Chinese listed companies. Our findings provide robust evidence that digital innovation positively influences ESG performance, with GAI technology adoption serving as a crucial mediating mechanism. This relationship is stronger for large firms, high-tech industries, and non-SOEs, suggesting important contextual contingencies. Furthermore, our analyses reveal that digital innovation and GAI technologies enhance ESG performance by improving operational efficiency, facilitating stakeholder engagement, and strengthening risk management.

Additionally, these findings offer several policy recommendations for promoting sustainable business practices through technological innovation. First, governments should develop integrated policy frameworks that simultaneously address digital transformation and sustainability objectives. This could involve incorporating sustainability criteria into digital innovation funding programs, establishing incentives for sustainable technology applications, and facilitating knowledge sharing on best practices at the intersection of digital innovation and ESG performance.

Second, regulatory agencies should develop standardized metrics and disclosure requirements for assessing firms' technological capabilities in addressing sustainability challenges. By enhancing transparency regarding the sustainability impacts of digital technologies, such measures would help investors identify firms that effectively leverage innovation for ESG improvements and incentivize other firms to follow suit.

Third, educational institutions and professional associations should develop training programs that integrate digital technology and sustainability knowledge. By nurturing talent with interdisciplinary expertise spanning both domains, such initiatives would address the skill gaps that currently hinder the effective utilization of digital technologies for sustainability purposes.

Fourth, international organizations should facilitate cross-border collaboration on sustainable technology development and adoption. Given the global nature of sustainability challenges and the rapid diffusion of digital technologies, coordinated international efforts could accelerate the development and deployment of technological solutions to pressing ESG issues.

Fifth, innovation ecosystems should be fostered that specifically target sustainability-oriented digital innovations. This could involve establishing specialized incubators, accelerators, and venture funds for sustainable technology startups, creating regulatory sandboxes for testing innovative solutions, and developing public-private partnerships focused on sustainable digital transformation.

In conclusion, our research highlights the transformative potential of digital innovation and GAI technologies in advancing corporate sustainability. By understanding and leveraging the synergies between technological advancement and ESG performance, firms can simultaneously enhance their competitiveness and contribute to addressing pressing environmental and social challenges. As digital technologies continue to evolve and diffuse, their strategic integration into sustainability initiatives will become increasingly important for creating long-term value for businesses and society.

**Funding:**

This research received no external funding"

**Institutional Review Board Statement:**

Not applicable

**Informed Consent Statement:**

Not applicable.

**Data Availability Statement:**

Not applicable.

**Acknowledgments:**

This is a short text to acknowledge the contributions of specific colleagues, institutions, or agencies that aided the efforts of the authors.

**Conflict of Interest:**

The authors declare no conflict of interest.

**References:**

Acemoglu, D., & Restrepo, P. (2019). Automation and new tasks: How technology displaces and reinstates labor. Journal of Economic Perspectives, 33(2), 3-30.

Agrawal, A., Gans, J., & Goldfarb, A. (2018). Prediction Machines: The Simple Economics of Artificial Intelligence. Harvard Business Review Press.

Bai, C., Quayson, M., & Sarkis, J. (2022). Digital business transformation and sustainable supply chain management: A systematic literature review. International Journal of Production Economics, 244, 108381.

Barney, J. (1991). Firm resources and sustained competitive advantage. Journal of Management, 17(1), 99-120.

Baron, R. M., & Kenny, D. A. (1986). The moderator-mediator variable distinction in social psychological research: Conceptual, strategic, and statistical considerations. Journal of Personality and Social Psychology, 51(6), 1173-1182.

Cui, J. (2025). The Impact of Absorptive Capacity, Organizational Creativity, Organizational Agility, and Organizational Resilience on Organizational Performance: Mediating Role of Digital Transformation. Organizational Creativity, Organizational Agility, and Organizational Resilience on Organizational Performance: Mediating Role of Digital Transformation (January 05, 2025).

Cui, J. (2025). The Explore of Knowledge Management Dynamic Capabilities, AI-Driven Knowledge Sharing, Knowledge-Based Organizational Support, and Organizational Learning on Job Performance: Evidence from Chinese Technological Companies. arXiv preprint arXiv:2501.02468.

Bharadwaj, A., El Sawy, O. A., Pavlou, P. A., & Venkatraman, N. (2013). Digital business strategy: Toward a next generation of insights. MIS Quarterly, 37(2), 471-482.

Brown, T. B., Mann, B., Ryder, N., Subbiah, M., Kaplan, J., Dhariwal, P., ... & Amodei, D. (2020). Language models are few-shot learners. Advances in Neural Information Processing Systems, 33, 1877-1901.

Brynjolfsson, E., & McAfee, A. (2017). The business of artificial intelligence. Harvard Business Review, 95(4), 3-11.

Brynjolfsson, E., Rock, D., & Syverson, C. (2019). Artificial intelligence and the modern productivity paradox: A clash of expectations and statistics. In The Economics of Artificial Intelligence: An Agenda (pp. 23-57). University of Chicago Press.

Bughin, J., Hazan, E., Ramaswamy, S., Chui, M., Allas, T., Dahlström, P., ... & Trench, M. (2018). Notes from the AI frontier: Modeling the impact of AI on the world economy. McKinsey Global Institute.

Cohen, W. M., & Levinthal, D. A. (1990). Absorptive capacity: A new perspective on learning and innovation. Administrative Science Quarterly, 35(1), 128-152.

Cowls, J., Tsamados, A., Taddeo, M., & Floridi, L. (2021). The AI gambit: Leveraging artificial intelligence to combat climate change—opportunities, challenges, and recommendations. AI & Society, 36, 1-25.

Damanpour, F., & Schneider, M. (2006). Phases of the adoption of innovation in organizations: Effects of environment, organization and top managers. British Journal of Management, 17(3), 215-236.

Zhou, L., & Cui, J. (2025). Dynamic Connectedness of Green Bond Markets in China and America: A R2 Decomposed Connectedness Approach. International Journal of Global Economics and Management, 6(2), 144-158.

Wan, Q., & Cui, J. (2024). Dynamic Evolutionary Game Analysis of How Fintech in Banking Mitigates Risks in Agricultural Supply Chain Finance. arXiv preprint arXiv:2411.07604.

Dwivedi, Y. K., Hughes, L., Ismagilova, E., Aarts, G., Coombs, C., Crick, T., ... & Williams, M. D. (2021). Artificial Intelligence (AI): Multidisciplinary perspectives on emerging challenges, opportunities, and agenda for research, practice and policy. International Journal of Information Management, 57, 101994.

Fichman, R. G., & Kemerer, C. F. (1997). The assimilation of software process innovations: An organizational learning perspective. Management Science, 43(10), 1345-1363.

Fiksel, J., Lambert, J. H., Artman, K. B., Harris, J. L., & Phifer, H. E. (2014). Environmental excellence: The new supply chain edge. Supply Chain Management Review, 18(1), 70-82.

George, G., Merrill, R. K., & Schillebeeckx, S. J. (2020). Digital sustainability and entrepreneurship: How digital innovations are helping tackle climate change and sustainable development. Entrepreneurship Theory and Practice, 44(6), 990-1000.

Goodfellow, I., Pouget-Abadie, J., Mirza, M., Xu, B., Warde-Farley, D., Ozair, S., ... & Bengio, Y. (2014). Generative adversarial nets. Advances in Neural Information Processing Systems, 27.

Hanelt, A., Bohnsack, R., Marz, D., & Antunes Marante, C. (2021). A systematic review of the literature on digital transformation: Insights and implications for strategy and organizational change. Journal of Management Studies, 58(5), 1159-1197.

Hart, S. L. (1995). A natural-resource-based view of the firm. Academy of Management Review, 20(4), 986-1014.

Hekkert, M. P., Suurs, R. A., Negro, S. O., Kuhlmann, S., & Smits, R. E. (2007). Functions of innovation systems: A new approach for analysing technological change. Technological Forecasting and Social Change, 74(4), 413-432.

Korinek, A., & Stiglitz, J. E. (2021). Artificial intelligence, globalization, and strategies for economic development. NBER Working Paper, (w28453).

Li, Y., Gong, M., Zhang, X. Y., & Koh, L. (2018). The impact of environmental, social, and governance disclosure on firm value: The role of CEO power. The British Accounting Review, 50(1), 60-75.

Liang, H., & Renneboog, L. (2017). On the foundations of corporate social responsibility. The Journal of Finance, 72(2), 853-910.

Lindgreen, A., Vallaster, C., Yousofzai, S., & Hirsch, B. (Eds.). (2019). Measuring and controlling sustainability: Spanning theory and practice. Routledge.

Nambisan, S., Lyytinen, K., Majchrzak, A., & Song, M. (2017). Digital innovation management: Reinventing innovation management research in a digital world. MIS Quarterly, 41(1), 223-238.

Nambisan, S., Wright, M., & Feldman, M. (2019). The digital transformation of innovation and entrepreneurship: Progress, challenges and key themes. Research Policy, 48(8), 103773.

Porter, M. E., & Kramer, M. R. (2011). Creating shared value. Harvard Business Review, 89(1/2), 62-77.

Preacher, K. J., & Hayes, A. F. (2008). Asymptotic and resampling strategies for assessing and comparing indirect effects in multiple mediator models. Behavior Research Methods, 40(3), 879-891.

Rahwan, I., Cebrian, M., Obradovich, N., Bongard, J., Bonnefon, J. F., Breazeal, C., ... & Wellman, M. (2019). Machine behaviour. Nature, 568(7753), 477-486.

Ransbotham, S., Kiron, D., Gerbert, P., & Reeves, M. (2017). Reshaping business with artificial intelligence: Closing the gap between ambition and action. MIT Sloan Management Review, 59(1).

Rogers, E. M. (2003). Diffusion of Innovations (5th ed.). Free Press.

Rolnick, D., Donti, P. L., Kaack, L. H., Kochanski, K., Lacoste, A., Sankaran, K., ... & Bengio, Y. (2022). Tackling climate change with machine learning. ACM Computing Surveys, 55(2), 1-96.

Russo, M. V., & Fouts, P. A. (1997). A resource-based perspective on corporate environmental performance and profitability. Academy of Management Journal, 40(3), 534-559.

Sarkis, J. (2021). Supply chain sustainability: Learning from the COVID-19 pandemic. International Journal of Operations & Production Management, 41(1), 63-73.

Schretzen, H., Wamba, S. F., Guillemette, M. G., & Omrani, H. (2021). The impact of artificial intelligence capabilities on corporate environmental, social, and governance (ESG) performance. Journal of Cleaner Production, 328, 129506.

Shrivastava, P. (1995). Environmental technologies and competitive advantage. Strategic Management Journal, 16(S1), 183-200.

Sobel, M. E. (1982). Asymptotic confidence intervals for indirect effects in structural equation models. Sociological Methodology, 13, 290-312.

Teece, D. J. (2007). Explicating dynamic capabilities: The nature and microfoundations of (sustainable) enterprise performance. Strategic Management Journal, 28(13), 1319-1350.

Teece, D. J., Pisano, G., & Shuen, A. (1997). Dynamic capabilities and strategic management. Strategic Management Journal, 18(7), 509-533.

Vial, G. (2019). Understanding digital transformation: A review and a research agenda. The Journal of Strategic Information Systems, 28(2), 118-144.

Vinuesa, R., Azizpour, H., Leite, I., Balaam, M., Dignum, V., Domisch, S., ... & Nerini, F. F. (2020). The role of artificial intelligence in achieving the Sustainable Development Goals. Nature Communications, 11(1), 1-10.

Whelan, T., & Fink, C. (2016). The comprehensive business case for sustainability. Harvard Business Review, 21(1), 1-12.

Wu, Z., Pan, S., Chen, F., Long, G., Zhang, C., & Philip, S. Y. (2019). A comprehensive survey on graph neural networks. IEEE Transactions on Neural Networks and Learning Systems, 32(1), 4-24.